\documentclass[usenatbib]{mn2e}
\usepackage[usenames]{color}
\input epsf
\usepackage{epsfig}
\usepackage{amsmath}
\usepackage{rotating}

\def\plotone#1{\centering \leavevmode
\epsfxsize=\columnwidth \epsfbox{#1}}

\def\spose#1{\hbox to 0pt{#1\hss}}

\def\approxlt{{\mathrel{\spose{\lower 3pt\hbox{$\sim$}}
        \raise 2.0pt\hbox{$<$}}}}
\def\approxgt{\mathrel{\spose{\lower 3pt\hbox{$\sim$}}
        \raise 2.0pt\hbox{$>$}}}

 \title[Width of X-ray lines in cooling flows]{Width of X-ray lines as
 a diagnostic of gas motions in cooling flows}

\author[P.~Rebusco et al.]{P.~Rebusco $^{1,5}$, E.~Churazov$^{1,2}$, R.~Sunyaev$^{1,2}$, H.~B\"ohringer$^{3}$,  W.~Forman$^{4}$ \\
$^1$ Max-Planck-Institut f\"ur Astrophysik, Karl-Schwarzschild-Strasse 1, 85741
Garching, Germany\\
$^2$ Space Research Institute (IKI), Profsoyuznaya 84/32, Moscow 117810, 
Russia\\
$^3$ MPI f\"{u}r Extraterrestrische Physik, P.O. Box 1603, 85740
Garching, Germany\\
$^4$ Harvard-Smithsonian Center for Astrophysics, 60 Garden St.,
Cambridge, MA 02138, USA\\
$^5$ Kavli Institute for Astrophysics and Space Research, MIT, Cambridge, MA 02139, USA
}

\pagerange{\pageref{firstpage}--\pageref{lastpage}}
\pubyear{2007}
\usepackage{subfigure}
\begin{document}

\maketitle
\label{firstpage}

\begin{abstract}

The dissipation of turbulent gas motions is one of the likely
mechanisms that has been proposed to heat the intracluster medium
(ICM) in the cores of clusters and groups of galaxies. We consider the
impact of gas motions on the width of the most prominent X-ray
emission lines. For heavy elements (like iron) the expected linewidth
is much larger than the width due to pure thermal broadening and the
contribution due to turbulent gas motions should be easily detected
with the new generation of X-ray micro-calorimeters, such as the
Spektr-RG calorimeter (SXC).  For instance in the Perseus cluster the
turbulent velocity required to balance radiative cooling (as derived
by Rebusco et al. 2006), would imply a width of the 6.7 keV Fe line of
10-20 eV, while the pure thermal broadening is $\sim$4 eV. 
The radial dependence of the linewidth is sensitive to i) the radial 
dependence of the velocity amplitude and ii) the "directionality" of the 
stochastic motions (e.g. isotropic turbulence or predominantly radial 
gas motions). If the width of several lines, characteristic for 
different gas temperatures, can be measured, then it should be possible 
to probe both the "directionality" and the amplitude of the gas 
motions.
Moreover a measurement of the width would put a lower limit on the amount of the  kinetic energy available for dissipation, giving a constraint on the ICM models.
\end{abstract}


\section{Introduction}

Clusters of galaxies are the largest gravitationally bound and nearly
virialized systems in the Universe. High resolution X-ray surveys have
revealed that the ICM (with temperatures in the range $\sim 2-10$ keV)
is not fully relaxed.  The substructures in the surface brightness and
temperature are observed both on large scales and in the cluster cores
indicating that the gas is not at rest. At the same time strong shocks
are rarely observed (see Markevitch \& Vikhlinin, 2007 for a review)
suggesting that the cluster gas motions are predominantly
subsonic. The shape of the emission lines as a source of information
on the ICM velocity distribution has been discussed in detail in
Sunyaev, Norman \& Bryan (2003) and Inogamov \& Sunyaev (2003).  These
papers mostly consider the gas motions induced by cluster mergers. In
the present paper we focus on the linewidth as a diagnostic of gas
motions in the cores of the galaxy clusters.

 In the cores of most clusters and many groups (e.g. Stewart et al
1984, Nulsen et al 1984, Edge \& Stewart 1991, Mulchaey et al 1993,
Fabian 1994), the radiative cooling time of the gas is short compared
to the age of the cluster and an external source of energy is needed
to avoid catastrophic cooling (e.g. David et al. 2001, Matsushita et al. 2002, Peterson
et al. 2003, Kaastra et al. 2004). The dissipation of turbulent gas
motions is one of the highly plausible sources of the required energy
(e.g. Pedlar et al. 1990, Churazov et al. 2002, Fabian et al. 2003a,
Chandran 2005). These motions can have distinctive effects on the
properties of the X-ray lines (e.g. Br\"{u}ggen, Hoeft \& Ruszkowski
2005). There are several indirect ways of probing the gas velocities,
i) using the resonant scattering of the brightest emission lines
(e.g. Gilfanov, Sunyaev \& Churazov, 1987, Churazov et al., 2004); ii)
considering the spreading of metals ejected by a central galaxy
through the ICM (Rebusco et al., 2005,2006, Graham et al., 2006,
Roediger et al., 2007); iii) using H$\alpha$ emitting filaments as
tracers of the gas motions (e.g. Fabian et al., 2003b, Hatch et al.,
2006). In particular in Rebusco et al. (2005,2006) the characteristic
spatial and velocity scales of the turbulent eddies were estimated
assuming that the dissipation of turbulent motions can generate enough
heat to prevent the gas from extreme cooling and that the same motions
provide moderate mixing of metals through the ICM. Using these
constraints we calculated the expected width of the brightest X-ray
lines for the brightest clusters (e.g., A426 - Perseus).

A micro-calorimeter with
energy resolution of less than $10$ eV and effective area of $\sim~200~$ cm$^2$ has been built for the Suzaku mission (Kelley et
al., 2007).  The effective areas of the upcoming calorimeters may
reach thousands and even tens of thousands of square cm in the next 10-15 years and
will have a spatial resolution between a few arcseconds and a few arcminutes
(see e.g. \verb#http://www.astro.isas.ac.jp/future/NeXT#,
\verb#http://www.rssd.esa.int/index.php?project=XEUS#,  \verb#http://constellation.gsfc.nasa.gov/# ).  

The Spektr-RG Calorimeter (SXC) (McCammon et al., 2007; Mitsuda et
al., 2007), proposed for the Spektr-RG Mission, has a peak effective
area of 300 cm$^2$ at $\sim 1.5$ keV, decreasing to about $30~$ cm$^2$
at $6.7$ keV and an energy resolution (goal) of $\sim 4$ eV.  As we
show in Section $3$, an instrument with these characteristics is well
suited to study the line broadening associated with turbulent gas
motions in the cores of the brightest galaxy clusters.

The structure of the paper is the following.  In section
\ref{sec:turb} we describe the line broadening under different
assumptions for the character of the gas motions; in section 3 we
consider the specific case of the Perseus cluster as an example and we
investigate the capabilities of SXC in this context. Our findings are
discussed and summarized in sections 4 and 5.

Throughout the paper when we mention turbulence, we mean all
types of macroscopic gas motions.

\section{The impact of turbulence on the linewidth}
\label{sec:turb}
There are several bright emission lines in the X-ray range, that are
characteristic of the ICM in galaxy clusters (Table
\ref{tab:lines}). The He-like iron line at 6.7 keV is the most
prominent feature above a few keV: it is especially bright in 
high temperature ($T_e > 4$ keV) clusters. For the lower temperature
clusters, and in particular for the cooling flow regions, L lines of
iron at $\sim$1 keV, lines of Si, Mg and O can be substantially
brighter than the 6.7 keV line. In Table \ref{tab:lines} we list a
sub-sample (incomplete) of these lines and provide an estimate of their
fluxes from the central $5'$ (radius) region of the Perseus
cluster. All line parameters were taken from the ATOMDB data base (Smith et al. 2001a,b).
\begin{table*}
\begin{center}
\begin{tabular}{c |c c c c c c c }	
\hline
\hline
Ion & Z   & u & l & E & $\epsilon_{\mbox{\small{peak}}}$ & T$_{\mbox{\small{peak}}}$ & F($<5'$)   \\
    &     &   &    & keV & phot cm$^3$ s$^{-1}$ & keV & $\times 10^{-3}$ erg cm$^{-2}$ s$^{-1}$ \\
\hline
FeXXV & $26$ & $1s2p~^1P_{1}$ & $1s^2~^1S_{0}$ & $6.700$   &  $4.56 \times 10^{-17}$   &  $5.42$ & $1.25$       \\  
FeXXIV & $26$ & $1s^23p~^2P_{3/2}$ & $1s^22s~^2S_{1/2}$ & $1.1675$   &  $9.92 \times 10^{-17}$   &  $1.74$ & $2.72$       \\  
SiXIV & $14$ & $2p~^2P_{3/2}$ & $1s~^2S_{1/2}$ & $2.006$   &  $5.02 \times 10^{-17}$ $^a$   &  $<1.50$ & $1.38$       \\  
MgXII & $12$ & $2p~^2P_{3/2}$ & $1s~^2S_{1/2}$ & $1.470$   &  $4.26 \times 10^{-17}$ $^a$   &  $<1.50$ & $1.17$       \\  
OVIII & $8$ & $2p~^2P_{3/2}$ & $1s~^2S_{1/2}$ & $0.650$   &  $2.26 \times 10^{-16}$ $^a$   &  $<1.50$ & $6.21$       \\             
\end{tabular}
\caption{Emission lines: (1) ion, (2) atomic number Z, (3)-(4)
respectively, upper and lower levels of the transition,  (5) line
energy in keV, (6) peak emissivity within the temperature range of interest
$1.5-10$ keV, (7) temperature (in the range $1.5-10$ keV) at which the
emissivity is the highest for a given line, (8) estimated flux of the line (phot
cm$^{-2}$ s$^{-1}$) for the central $5'$ region of the Perseus
cluster.
$(a)$ Line emissivity at $T=$1.5 keV. The true peak
emissivity in this line is at a temperature below 1.5 keV.
}
\label{tab:lines}
\end{center}
\end{table*}

Let us first assume that i) the cluster is spherically symmetric, ii)
the characteristic correlation length of the velocity field is much smaller
than the characteristic dimensions of the system and iii) at each
location the emission line profile can be approximated by a Gaussian:
\begin{eqnarray}
p(\nu-\nu_0)=\frac{1}{\sqrt{2 \pi}}\frac{1}{\sigma_\nu} \exp{\left(- \frac{1}{2}\frac{(\nu-\nu_0)^2}{\sigma_\nu^2} \right)},
\end{eqnarray}
where $\nu_0$ is the frequency of the transition and and $\nu$ is the observed frequency,
$\sigma_\nu=\sigma_\nu(R)$ is the width of the line, which is a
function of the distance R from the cluster center.  The
spectral surface brightness in a line at a given projected distance ($x$)
from the center of the cluster is given by the following integral
along the line of sight ($l$):
\begin{eqnarray}
I(\nu ,x) = \int_{-\infty}^\infty n_e^2~a~
              \epsilon_{\nu_0}~ p(\nu-\nu_0)~dl, 
\label{eq:int}
\end{eqnarray}
where $n_e=n_e(R)$ is the electron density, $a=a(R)$ the element
abundance, $\epsilon_{\nu_0}=\epsilon_{\nu_0}\left(T_e(R)\right)$ 
the plasma emissivity in a given line, which is a function of the
plasma temperature $T_e(R)$. In the above expression
$R=\sqrt{x^2+l^2}$. The line broadening is decomposed into two components:
\begin{eqnarray}
\sigma^2_\nu=\sigma^2_{thermal}+\sigma^2_{turb},
\label{eq:sigma}
\end{eqnarray}
where
\begin{eqnarray}
&\sigma_{thermal}=\nu_0\frac{\sqrt{\frac{kT_e}{A m_p}}}{c}\nonumber \\
&\sigma_{turb}=\nu_0\frac{v_{\parallel}}{c},
\label{eq:sigmas}
\end{eqnarray}
where $k$ is the Boltzmann constant, $m_p$ the proton mass, $A$ the
atomic weight of the element, $c$ the speed of light and $v_{\parallel}$ the line of sight component of the turbulent velocity. For the case of 
isotropic turbulence, $v^2_{\parallel}=v^2_{turb}/3$, where
$v_{turb}$ is the root-mean-square of the 3-dimensional turbulent velocity. A comparison
of  thermal and turbulent broadenings in a 4 keV plasma is given in
Table \ref{tab:fwhm}. Note that even for the iron line at 6.7 keV, the
natural width of the line due to radiative decay is only $\sim$0.2 eV
(and it is much lower for the other transitions in Table \ref{tab:fwhm}). We
therefore neglect the natural width of all the lines throughout this paper.

\begin{figure}
\plotone{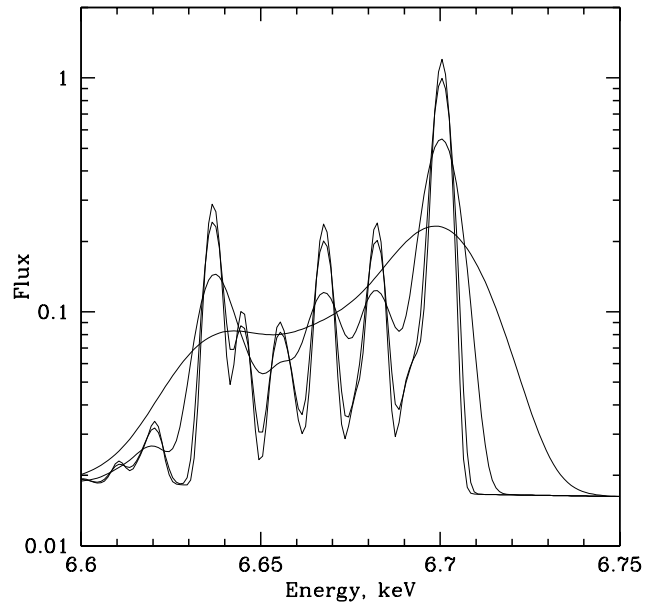}
\caption{6.7 keV line complex for a $T=$4 keV plasma. The four spectra correspond to thermal broadening and turbulent broadening with
$v_{turb}$=0, 100, 300 and 900 km/s. For comparison the sound speed in
a 4 keV plasma is $\sim$1000 km/s.}
\label{fig:spec}
\end{figure}

\begin{table*}
\begin{center}
\begin{tabular}{c |c c c c c }	
\hline
\hline
Ion, E(keV)& $v_{turb}$(km/s)    & FWHM$_{thermal}$(eV) & FWHM$_{turb}$(eV) & $\frac{FWHM_{total}}{FWHM_{thermal}}$  \\
\hline
FeXXV, $6.700$   &  $100$   &  $4.35$ & $3.04$ & $1.22$       \\  
            &  $300$   &  $4.35$ & $9.12$ & $2.30$       \\
            &  $900$   &  $4.35$ & $27.35$ & $6.31$       \\
FeXXIV, $1.170$   &  $100$   &  $0.76$ & $0.53$ & $1.22$       \\  
            &  $300$   &  $0.76$ & $1.59$ & $2.32$       \\
            &  $900$   &  $0.76$ & $4.78$ & $6.36$       \\
SiXIV, $2.006$   &  $100$   &  $1.84$ & $0.91$ & $1.12$       \\  
            &  $300$   &  $1.84$ & $2.73$ & $1.79$       \\
            &  $900$   &  $1.84$ & $8.19$ & $4.55$       \\
MgXII, $1.470$   &  $100$   &  $1.46$ & $0.67$ & $1.10$       \\  
            &  $300$   &  $1.46$ & $2.00$ & $1.70$       \\
            &  $900$   &  $1.46$ & $6.00$ & $4.23$       \\
OVIII, $0.650$   &  $100$   &  $0.79$ & $0.29$ & $1.07$       \\  
            &  $300$   &  $0.79$ & $0.88$ & $1.50$       \\
            &  $900$   &  $0.79$ & $2.65$ & $3.50$       \\
\hline
\end{tabular}
\caption{Thermal and turbulent broadening in a 4 keV plasma: (1) line
energy in keV, (2) turbulent velocity in km/s, (3) FWHM due to pure thermal broadening, (4) FWHM due to isotropic turbulent broadening alone,
(5) ratio of the total FWHM to pure thermal broadening.}
\label{tab:fwhm}
\end{center}
\end{table*}
It is obvious from equation \ref{eq:sigmas} that the heaviest elements are the best
probes of gas motions, since more massive
nuclei have smaller thermal line broadening and, therefore, the
turbulent broadening can dominate the thermal broadening. Indeed the ratio
of the e.g. iron thermal velocity to the proton thermal velocity is
small ($\sim (m_p/A m_p)^{1/2}=0.13$). Thus even for small velocities
the turbulent broadening will exceed the thermal broadening (Table
\ref{tab:fwhm}, Figure $1$). For an instrument with a given spectral resolution
the energy of the line also plays an important role, since the
width of the line simply scales with the line energy. This makes the
iron 6.7 keV line the best probe of the turbulent ICM, as long as the
instrument has a substantial effective area at 6-7 keV.

The above assumption of a Gaussian line shape due to turbulent motions is of
course only an approximation (see Inogamov \& Sunyaev 2003 for a discussion). However it is fully sufficient to assess the
detectability of the turbulent broadening. In  fully developed
turbulence an entire range of scales is produced through an energy
cascade (e.g.  Richardson 1922, Tennekes \& Lumley 1972, 
Lesieur 1997, Mathieu 2000, Davidson 2004), which is
characterized by large scale energy-containing motions and a small
dissipation scale. In our simplified treatment we assume that the 
$v_{turb}$ entering all our equations can be obtained from the kinetic
energy of the turbulent motions:
$\epsilon_K=\rho\frac{v^2_{turb}}{2}$, where $\rho$ is the gas
density. Since the integration of equation \ref{eq:int} may result in a
line shape different from a pure Gaussian we use below an
``effective'' full width half maximum (eFWHM), which is
defined as the interval of energies centered at the line energy and
containing $76\%$ of the line flux.

\subsection{Isotropic, radial and tangential motions}
\label{sec:iso}
\begin{figure}
\plotone{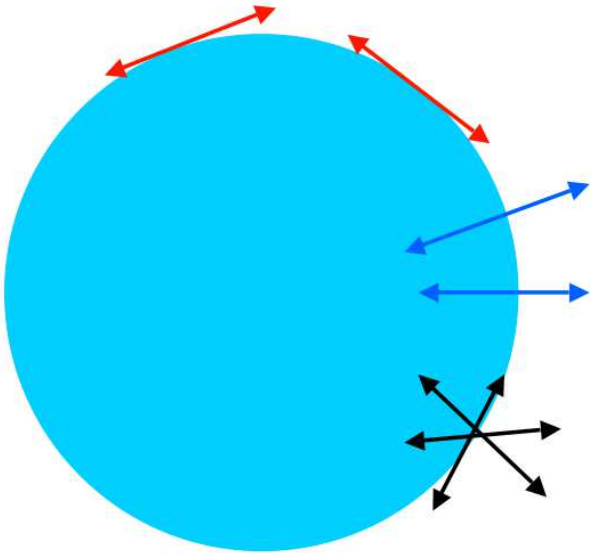}
\caption{Sketch of isotropic (lower right, black arrows), radial
  (right, blue arrows) and tangential motions (top, red arrows). }
\label{fig:sketch}
\end{figure}
There is indirect evidence that the gas in cluster centers
is involved in some sort of motion. H$\alpha$ emitting filaments, that are thought to be drawn behind buoyant gas bubbles, have been used to trace the ICM velocity field (e.g. Hatch et al. 2006, 2007, Salom\'e et al. 2007). However neither the characteristic
velocity scales nor the characteristic patterns of the hot gas motions have
been measured in a more direct way so far. One can identify two major sources of
turbulent gas motions in the cores: cluster mergers and the  action
of a central active galactic nucleus (AGN). Each process
may produce a velocity field different in intensity and directionality.
Let us consider the extreme cases  of  isotropic, pure radial and
pure tangential velocity fields as illustrated in Figure
\ref{fig:sketch}. If one fixes the total kinetic energy in the motions,
then the line of sight component of the velocity can be written as:
\begin{equation}
v_{\parallel}^2 =
\begin{cases}
\frac{v_{turb}^2}{3} & \text{isotropic,}\\
v_{turb}^2 \frac{l^2}{R^2} & \text{radial,}\\
\frac{v_{turb}^2}{2}\frac{x^2}{R^2} & \text{tangential.}
\end{cases}
\label{eq:vel}
\end{equation}
The first case (isotropic gas motions)  can occur for
a wide variety of driving mechanisms, ranging from mergers to the
motions caused by rising buoyant bubbles of relativistic plasma or to
convection driven by a mixture of thermal plasma and cosmic rays
(Chandran 2005). Pure radial gas motions would naturally appear if  energy generated by a central AGN goes into weak shocks and
sound waves (e.g. Fabian et al. 2003, Forman et al. 2005, 2006, Fabian et al. 2006) which propagate through the ICM almost
radially from the central source.

\begin{figure}
\plotone{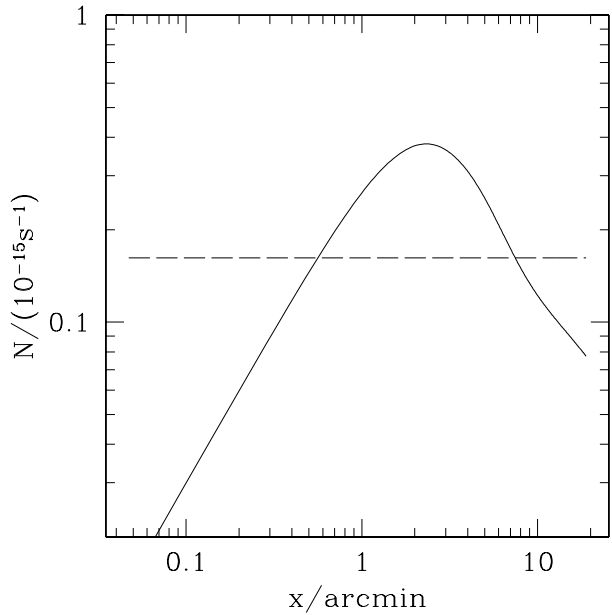}
\caption{Brunt-V\"ais\"al\"a frequency $N$ (solid line) for the Perseus cluster
in comparison with the characteristic frequency of turbulent motions
$v_{turb}/l$ for $v_{turb}=~100~~$ km s$^{-1}$ and $l=20$ kpc (dashed line). If gravity waves are excited, then they are trapped in the region where the turbulent frequency is smaller than the Brunt-V\"ais\"al\"a frequency. }
\label{fig:bv}
\end{figure}

 The case of pure tangential
motions seems to be less natural and it is included mostly for
completeness. We note however that in the stratified atmosphere of
clusters the characteristic frequencies of the turbulent motions can
be smaller than the Brunt-V\"ais\"al\"a frequency:
\begin{equation}
N^2=g\left(\frac{1}{\gamma~P}\frac{dP}{dr}-\frac{1}{\rho}\frac{d\rho}{dr} \right),
\end{equation}
where $g$ is the gravitational acceleration due to the dark matter
potential, $\rho$ is the gas density, $P$ is its pressure and
$\gamma=5/3$ is the adiabatic index. In Figure \ref{fig:bv} we plot the 
Brunt-V\"ais\"al\"a frequency, evaluated for the Perseus cluster, and
compare it with the characteristic frequency of turbulent motions
$v_{turb}/l$ for $v_{turb}=~100~$ km s$^{-1}$ and $l=20$ kpc. 
In the range where the turbulent frequency is smaller than the local Brunt-V\"ais\"al\"a frequency, the gas motions may excite gravity waves
(e.g. Churazov et al., 2001, 2002, Omma et al., 2004). 
These waves can propagate and transport energy from a localized patch of turbulent ICM within this region (e.g. in the case of Perseus within a radius of $\sim 10$ arcmin). Such a
process is well known in atmospheric science and oceanography and it
may lead to the formation of non-propagating pancake-shaped vortices
(e.g. Riley \& Lelong 2000).   Therefore, one can imagine that patches of
predominantly 2-dimensional (tangential) vortices can be created with
this mechanism.

\begin{figure}
\plotone{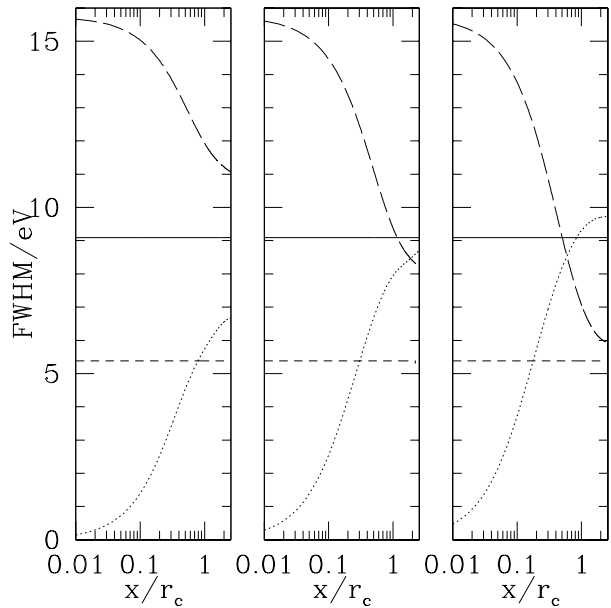}
\caption{6.7 keV linewidth as a function of the projected radius, for
isotropic (solid), tangential (dotted) and radial (long-dashed)
turbulence. The profile is obtained by integrating equation
\ref{eq:int}, as explained in section \ref{sec:turb}. In the three
panels the gas density profile is different: from left to right
$\beta=0.3,0.5,0.9$.  The thermal broadening is neglected, but its
magnitude in a plasma at $6$ keV is shown for reference (short-dashed
line).}
\label{fig:beta}
\end{figure}

We now calculate (Figure \ref{fig:beta}) the expected width of the iron
6.7 keV line for each of the three limiting cases given in
equation \ref{eq:vel}, for a model cluster with the density parameterized by
a simple $\beta$ model: $n_e=n_e(0)\left[1+(r/r_c)^2 \right]^{-3/2
\beta}$, where $r_c$ is the core radius of the cluster and $\beta=0.3,0.5,0.9$. The gas
temperature and the abundance of iron are assumed to be constant.  The
velocity $v_{turb}$ is fixed to 300 $\rm km~s^{-1}$. The thermal broadening
(first term in equation \ref{eq:sigma}) is omitted in the calculation of the
linewidth, but the level of thermal broadening for a 6 keV plasma is
shown for comparison (dashed line). The three plots in
Figure \ref{fig:beta} correspond to $\beta=0.3,0.5,0.9$ from left to
right, respectively. As expected from equation \ref{eq:vel}, isotropic
turbulence produces a linewidth independent of the projected distance
from the cluster center. Pure radial gas motions would result in a line whose width is peaked towards the center of the cluster, while pure
tangential motions produce the broadest line outside the cluster
core. For $x\gg r_c$, the curve approaches asymptotic values
(different for each velocity pattern), that are functions of
$\beta$ only. For large $\beta$ the cluster is more ``compact'' and
the gradients of the width of the line with radius become stronger, while
for low $\beta$ the change of the linewidth is more gradual.

\subsection{Radially dependent velocity amplitude}
\label{subsec:central}

It is quite plausible that the characteristic turbulent velocity scale varies with the distance from the cluster center. 
If gas motions are driven by the outflows of a central AGN, then
their amplitude will likely decrease for large radii, since the energy
would be spread over larger masses of gas and it would be partly
dissipated at smaller radii. For example, the
characteristic amplitude of spherical sound waves propagating through
a declining density profile decreases as $v\propto
\left(\rho~R^2\right)^{-1/2}$. For  $\rho(R)\propto 1/R$ the
velocity varies as $v\propto \left(R\right)^{-1/2}$. If, furthermore,
dissipation is taking place then the amplitude will decline more
quickly, as suggested by Fabian et al. (2003) and studied in three-dimensional viscous simulations by Ruszkowski et al. (2004). For rising buoyant bubbles of relativistic plasma the
terminal velocity scales as $v\propto v_K\sqrt{r_b/R}$ (e.g. Churazov
et al., 2001), where $v_K$ is the Keplerian velocity at radius $R$ and
$r_b$ is the characteristic size of the bubble. The adiabatic
expansion of the bubble leads to a slow change of the bubble size
$r_b\propto P^{-1/4}$ (assuming that the adiabatic index of the medium
inside the bubble is 4/3). Thus the characteristic bubble velocity is
$v\propto v_K R^{-1/2} P^{-1/8}$ - a decreasing function of the radius
for plausible $v_K(R)$ and $P(R)$. If the bubble breaks down into
smaller bubbles, then the terminal velocity will decrease even further
with the radius. 

Alternatively, if the turbulence is driven by shocks originating from
minor mergers, then it is possible that the velocity amplitude will
instead decline towards the center, where the gas density and the
thermal gas pressure are the highest. For example, for a plane sound
wave propagating into a region of increasing density the
characteristic velocity is $v\propto \rho^{-1/2}$, i.e. $v\propto
R^{1/2}$ if $\rho\propto 1/R$. Here we have ignored both the possible
reflection of the sound waves by a steep density gradient (important
for long wavelength perturbation) and the focusing of sound waves due
to decreasing sound speed towards the center (Pringle, 1989). In
reality the effect of mergers on the cluster core is much more
complicated and the above estimate can at best be considered as
indicative.

\begin{figure}
\plotone{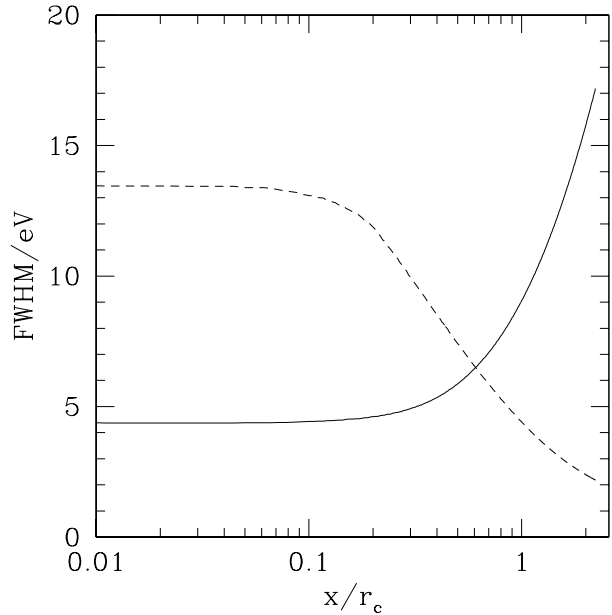}
\caption{6.7 keV linewidth as a function of the projected radius, for
isotropic turbulence when the velocity scale follows the law
$v(R)=v_0\left(\frac{R}{r_c}\right)^\alpha$, with $\alpha=\pm 1$.  In
each case the velocity scale was assumed to saturate at $1000$ km/s.
The solid line is for turbulent velocity increasing with radius, the
dashed line for turbulent velocity declining with radius.  For clarity
the thermal broadening is neglected. }
\label{fig:raising}
\end{figure}

For illustration we model two possibilities (decreasing or increasing
turbulence) by considering
$v(R)=v_0\left(\frac{R}{r_c}\right)^\alpha$, with $\alpha=\pm 1$ and
$v_0=200$ km/s. We further assume that when $v(R)>1000$ km/s, the
velocity saturates at $1000$ km/s. The resulting linewidth profiles
are shown in Figure \ref{fig:raising}. Clearly the radial dependence of
the velocity amplitude can strongly affect the behavior of the
linewidth as a function of projected distance.

\section{An example: the Perseus Cluster}
\label{sec:perseus}
We now calculate the expected width of the 6.7 keV line by using the Perseus
cluster as an example. Perseus (A 426) is the brightest nearby X-ray
cluster and it is one of the best-studied cool core clusters, together
with M87 and Centaurus.  It hosts in its core a luminous elliptical
galaxy NGC 1275, containing a bright radio source (3C 84). In the core
region a complex substructure is seen in X-ray
temperature, X-ray surface brightness and optical light distributions. Such substructure includes holes in the
X-ray images due to bubbles of relativistic plasma (B\"ohringer et
al., 1993, Fabian et al. 2000), quasi-spherical ripples (Fabian et
al. 2003a) and optical H$\alpha$ filaments (Fabian et al. 2003b, Hatch et
al. 2005).

\begin{figure}
\plotone{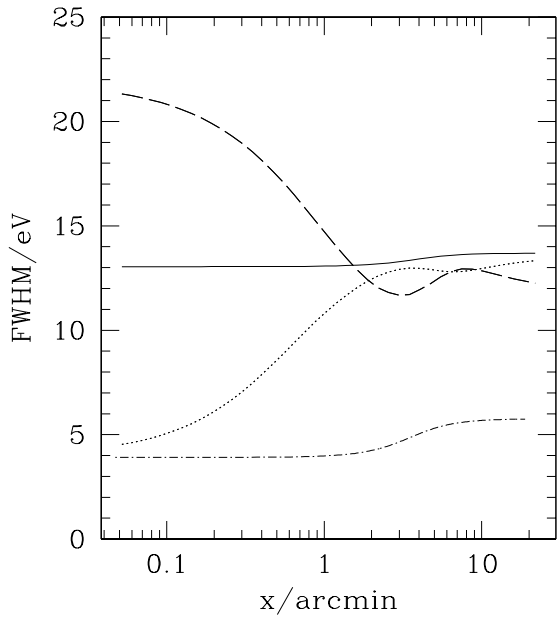}
\caption{6.7 keV linewidth as a function of the projected
radius in the Perseus cluster for isotropic (solid), radial
(long-dashed) and tangential (dotted) motions. The characteristic
turbulent velocity $v_{turb}=410~{\rm km/s}$ was assumed to be
constant with radius.  The thermal broadening (shown with the
short-dashed line) was included in the calculation of the linewidth.}
\label{fig:pdir}
\end{figure}
\begin{figure}
\plotone{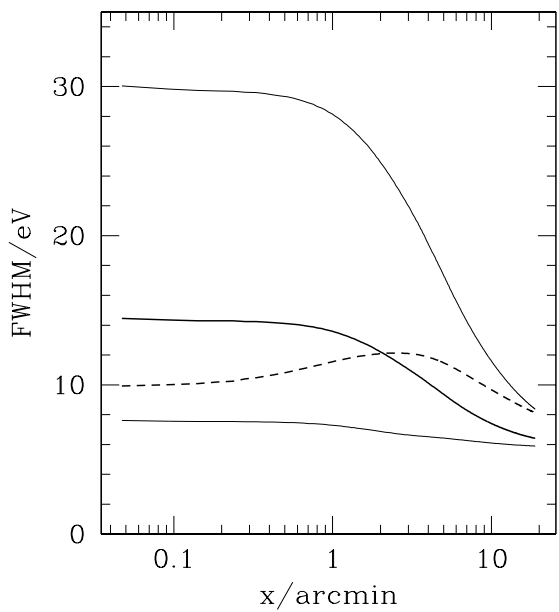}
\caption{Expected radial dependence of the 6.7 keV linewidth in the
  Perseus cluster for several illustrative cases. For all curves we
  assume a balance between gas cooling losses and turbulent heating at
  each radius according to eq.\ref{eq:bal}. The solid lines correspond to
  $l=$ 2, 20, 200 kpc (from the bottom to the top respectively), independently of the radius. The dashed line
  corresponds to the $l=0.3 R$. }
\label{fig:heat}
\end{figure}


%


In our estimates of the expected linewidth in Perseus, we assume that the
gas cooling losses are compensated by the dissipation of the turbulent
motions at all radii. This is of course a strong and not fully
justified assumption, but it provides clear predictions if the turbulent heating is indeed important in cluster
cores. Thus one can write:
\begin{eqnarray}
C~\rho~v_{turb}^3/l\approx n^2_e \Lambda(T),
\label{eq:bal}
\end{eqnarray}
where $\Lambda(T)$ is the gas cooling function and $C$ is a
dimensionless constant of the order of unity (see e.g. Dennis \& Chandran,
2005), which depends on the character of the turbulent motions. Since
the electron density and temperature are known from observations
(e.g. Churazov et al., 2003) one can estimate $v_{turb}^3/l$ from equation
\ref{eq:bal}. As the linewidth depends explicitly on $v_{turb}$ one has
to determine (or assume) some constraints on $l$. For instance,
Rebusco et al. (2005, 2006) and Graham et al. (2006) considered
the turbulent spreading through the ICM of metals produced by the brightest cluster
galaxy.  Treating this process in a diffusion
approximation one can estimate the effective diffusion coefficient
$D$. In the Perseus cluster $D\sim 2~\times~10^{29}~\rm cm^2/s$ (Rebusco et
al. 2006). One can then cast $D$ in the form $D=C'~v_{turb}~l$, where $C'$
is a dimensionless constant of the order of unity (see Dennis \&
Chandran 2005 for a compilation of values of $C$ and $C'$). Combining
equation \ref{eq:bal} (at some radius) and the expression for
the diffusion coefficient, both $v_{turb}$ and $l$ can be measured. In Perseus the
characteristic values were found to be $l\sim$ 20 kpc and $v_{turb}\sim$ 410
km/s (Rebusco et al. 2006). If we adopt $v_{turb}=$410 km/s as the characteristic velocity of turbulent
motions at every radius in Perseus then, depending on the
directionality of the turbulence (see Figure \ref{fig:pdir}), the 
expected width of the 6.7 keV line in Perseus should be in the range
10-20 eV. Of course the above values of $l$ and $v_{turb}$ are only order of magnitude estimates. From equation \ref{eq:bal} it is clear that $v_{turb}\propto
\left(n_e~l~\Lambda(T)\right)^{1/3}$.  If $l={\rm const}$ then $v_{turb}\propto
n_e^{1/3}\propto R^{-1/3}$ (when $n_e\propto 1/R$). In Figure \ref{fig:heat} we
calculate the linewidth for several illustrative cases: three solid
lines with different thickness correspond to $l=$2, 20, 200 kpc (at
all radii). This spread of two orders of magnitude in $l$ results in a
factor of 3 change of the linewidth. If instead we follow the arguments of
Dennis \& Chandran (2005) and set $l\sim\alpha R$, where $\alpha=0.3$,
then $v_{turb}\approx$ const when $n_e\propto 1/R$. The corresponding curve is
shown in Figure \ref{fig:heat} by the dashed line.


\begin{figure}
\plotone{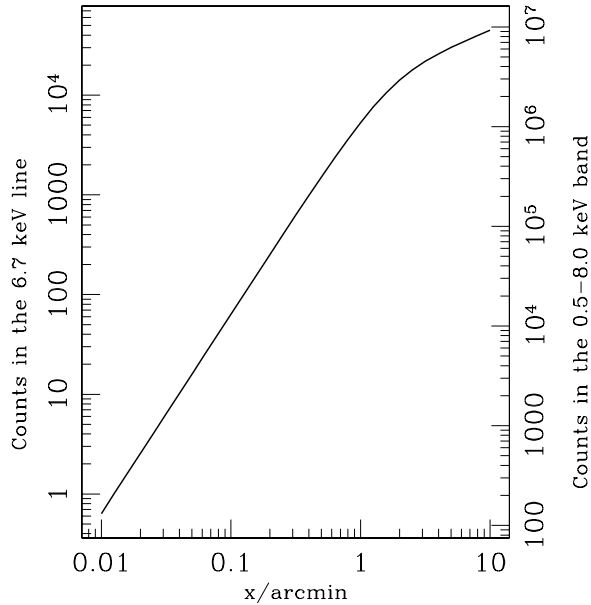}
\caption{The estimated total photon counts in the 6.7 keV line coming from a circle enclosed in a given projected radius from
  the Perseus cluster. The assumed duration of the observation is 1 Msec. The counts in the 0.5-8.0 keV band are shown for reference on the right axis.}
\label{fig:flux}
\end{figure}

\begin{figure}
\centering \leavevmode
\epsfig{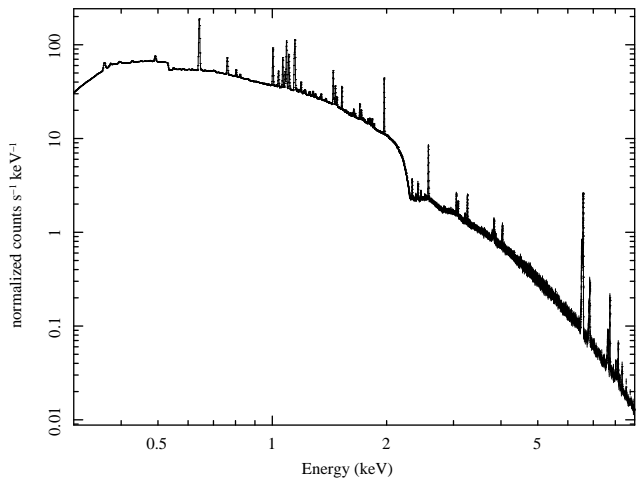}
\caption{Simulated spectrum of the Perseus cluster core within a radius of $6'$, using the properties of the proposed SXC. The spectrum is derived for $v_{\parallel}=236.714$ km/s, $T=4$ keV and an exposure time of $10^6$ s.}
\label{fig:xspec}
\end{figure}

\begin{figure}
\plotone{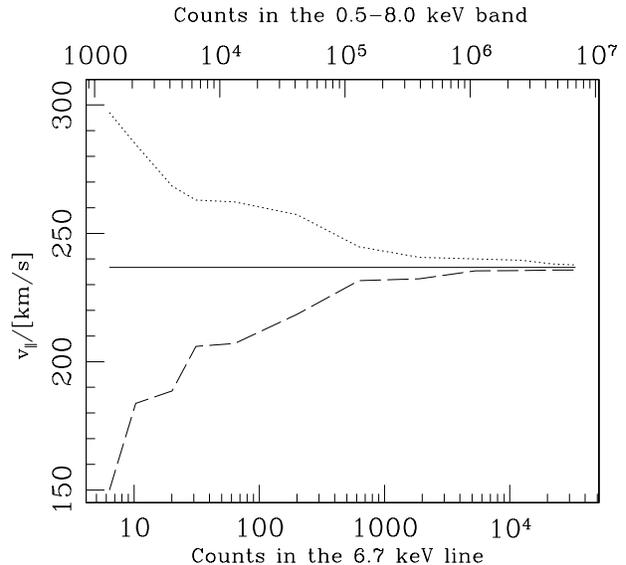}
\caption{Uncertainty in the measured 6.7 keV line broadening as a function of the number of counts (in the line and in the 0.5-8.0 keV band). The solid line shows the true broadening ($237$ km/s), while the long-dashed and dotted lines indicate the level of $1$ sigma confidence for the fitted velocity. The comparison with Figure 8 suggests that for a circular region with a radius of $0.5'$ in the core of Perseus, the turbulent velocity broadening can be recovered with an uncertainty of about $5$ km/s (for a 1 Msec long observation).}
\label{fig:fake}
\end{figure}

In order to make some estimates we consider the proposed SXC with an
effective area which can be provided by the
current generation of X-ray mirrors with a rather short focal length
of less than $2$ meters. At $6.7$ keV the effective area is $\sim30$
cm$^2$. We adopt a spectral and spatial resolution of $4$ eV and $1$
arcmin respectively. The design goal for SXC is 4 eV resolution, a
field of view of $11 \times 11$ arcmin square and pixels of $1.8$
arcmin square. We want to check how accurately a
micro-calorimeter with such parameters will measure the turbulent
broadening in the Perseus cluster. The expected number of photon
counts in the iron complex in a 1 Msec observation is plotted in
Figure \ref{fig:flux}. Note that, although we focus on the line
broadening of the $6.7$ keV line, in what follows we use all the
photon counts in the 0.5-8.0 keV band, in order to obtain better fits
(for reference the number of counts due to the $6.7$ keV line accounts
for $\sim 0.48 \%$ of the total number of counts in the 0.5-8.0 keV
band).  We simulated the observed spectra using XSPEC, specifically
the \textit{bapec} model, which takes into account isotropic
broadening. For $v_{turb}^{iso}=410$ km/s, the component of the
velocity along the line of sight is about $237$ km/s. An example of
the simulated spectrum (for a 1 Msec long observation) is shown in
Figure \ref{fig:xspec}.
The uncertainty in the measured 6.7 keV line broadening as a function of the number of counts is plotted in Figure \ref{fig:fake}.
The solid line shows the true broadening ($237$ km/s), while the long-dashed and dotted lines indicate the level of $1$ sigma confidence for the fitted velocity. The comparison with Figure 8 suggests that already within $0.5$ arcmin there are enough photons to measure the
turbulent velocity with an uncertainty of $\sim~\pm~5$ km/s.
Figure \ref{fig:fake} assumes a plasma temperature of $4$ keV. We
repeated the simulations also for higher temperatures and assuming a
2-temperature plasma. In both cases, the turbulent velocity can be
recovered with an uncertainty below $15$ km/s. In order to test the
capabilities of our fiducial micro-calorimeter, we also calculated the
accuracy of the fits within each single square of $1.8 \times 1.8$ arcmin.
Even in the outer parts, the resolution is high enough to measure the
turbulent velocity with $1$ sigma confidence below $10$
km/s. Therefore the spatial dependence of the turbulent velocity could
be directly measured and this would give clear indications about the
origin of the turbulent gas motions (see Section $2$).

\section{Linewidth versus shift of the line centroid}
\label{sec:centroid}
All the results obtained in the previous sections are based on the
assumption that the characteristic size of the turbulent eddies $l$ is
much smaller than the characteristic length of the line of sight. 
\begin{figure}
\plotone{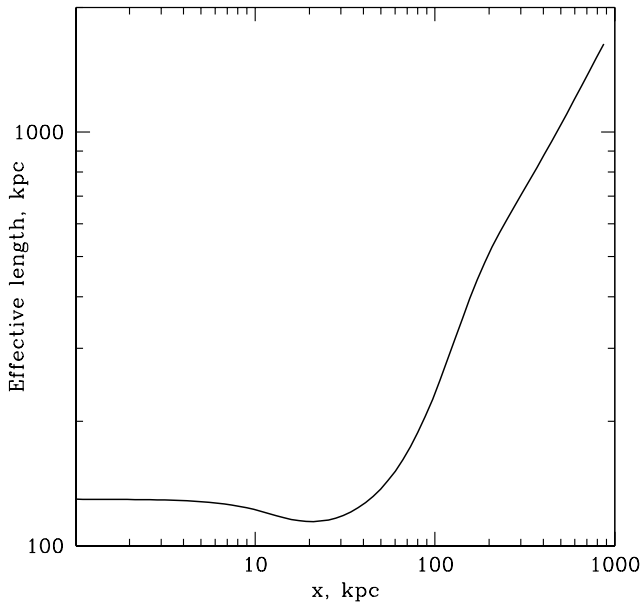}
\caption{Effective length of the region along the line of sight which
provides 75\% of the total surface brightness in the 6.7 keV line at a
given projected distance from the center of the Perseus
cluster. The ``dip'' around 20-30 kpc is due to the complicated structure
in the assumed radial abundance profile (e.g. Schmidt et al. 2002, Churazov et al. 2003).}
\label{fig:el}
\end{figure}

In practice there are at least two complementary ways of deriving
information on the turbulent gas motions with micro-calorimeters. One
can use i) the variations of the line centroid with position and/or
ii) the width/shape of the line. Assuming that the
resolution of the instrument is substantially better than the width
 of the observed line  (e.g. see Fig. \ref{fig:heat}) the accuracy of
determination of the line characteristics (in the case of a strong single
line with negligible continuum, which corresponds to the most
optimistic scenario) is:
\begin{eqnarray}
\sigma_{centroid}&\approx&\frac{{\rm FWHM}_{line}}{2
\sqrt{2~\ln{2}} ~ \sqrt{N_{counts}}} \\
\sigma_{FWHM}&\approx&\frac{{\rm FWHM}_{line}}{\sqrt{2}~\sqrt{N_{counts}}}, 
\end{eqnarray}
where ${\rm FWHM}_{line}$ is the width of the line, $N_{counts}$ is
the number of counts in the line, $\sigma_{centroid}$ and
$\sigma_{FWHM}$ are the 1$\sigma$ errors in the centroid and width of the
line.\\
The centroid variation method works best only if the characteristic size of the eddies is not too
small. Indeed, in practice one observes the spectrum coming from a
selected region of the cluster with characteristic sizes $X$ (in
projection) and $L$ (along the line of sight). If $X\gg l$ and $L\gg
l$ then $N_{eddies}\sim X^2L/l^3$ independent eddies are seen by the
instrument at the same time. Assuming that each eddy produces a
centroid shift of the order of $v_{\parallel}$, then the observed
centroid shift will be $\sim v_{\parallel}/\sqrt{N_{eddies}}\propto
X^{-1}L^{-1/2}$. Increasing $X$ would increase the number of line
photons the instrument detects during observations, but this would
simultaneously decrease the amplitude of the centroid shift. Obviously
these two effects cancel each other and increasing $X$ beyond the
characteristic eddy size $l$ does not improve the detectability of the
centroid variations. Making $X$ smaller than $l$ would just make the
number of line photons smaller: the amplitude of the centroid
variation should not vary much. Of course studying the centroid
variations on spatial scales larger and smaller than $l$ will be a
very useful test of the distribution of characteristic eddy sizes, but
for the principal detection of the gas turbulent motions, the choice
of $X\approx l$ is obviously the optimal one. Thus the question of
prime importance is whether the instrument has an effective area large
enough to collect photons from a region with size $X\sim l$
(e.g. sufficient to detect the centroid variations). Of course,
averaging along the line of sight remains an unavoidable effect
independent of the choice of the region size. In Figure \ref{fig:el}
we plot the effective length $L$ of the region along the line of sight
which provides 75\% of the total surface brightness in the 6.7 keV
line, at a given projected distance from the center of the Perseus
cluster. It is clear from this figure that, in the core of the Perseus
cluster, $L$ is of the order of 130 kpc and it increases linearly
outside the core. The non-monotonic behavior of the curve at 20-30 kpc
from the center is caused by the complicated iron abundance profile
(see e.g. Schmidt et al. 2002, Churazov et al. 2003). Thus, for
the optimal size of the observed region, the typical amplitude of the
centroid variations will be $v_{\parallel}/\sqrt{L/l}$. For estimating
the centroid variations, we adopt $v_{turb}\sim$ 400 km/s (i.e. in the
isotropic case $v_\parallel\sim$ 237 km/s) and $l\sim$ 20 kpc (Rebusco
et al., 2006). In the Perseus cluster a single $1.8'
\times 1.8'$ SXC pixel, corresponds to a $41 \times 41$ kpc region
pixel, and an effective length along the line of sight of $\sim 100$ kpc (see
Fig.\ref{fig:el}). Thus one can expect $\approx 41*41*100/20^{3}=21$ eddies
(assuming a $20^{3}~{\rm kpc^3}$ volume per eddy) and the typical centroid
variations on these spatial scales are of the order of 50 km/s. Such a
shift in velocity corresponds to a $\sigma_{line}\sim$ 1.1 eV shift in the line
energy from pixel to pixel for the 6.7 keV line. 
From eq. 8  it follows that such variations should be
detectable with $\sim 1\sigma$ significance in SXC pixels 
with $\sim 100$ counts (for the 6.7 keV line with FWHM of 13 eV - as in
Fig. 6). As one can see in Fig. 8 the number of photons in the central pixel for a 1 Msec observation of the Perseus cluster is much higher (of the order of $10^4$): hence the centroid variation can be measured with great accuracy.\\
The detection of the line broadening has no limitation from the point
of view of the region size and the whole cooling flow region can be
probed at once. According to eq. 9 the 1$\sigma$ error on the width of
the line is $\sim 0.3$ eV (for the 6.7 keV line with FWHM of 13 eV
and number of detected photons $N_{counts}=1000$). This corresponds to
a change in the characteristic velocity scale of $\sim 13$ km/s. In a 1
Msec observation such broadening can be measured in each of the $1.8'
\times 1.8'$ pixels of SXC up to a distance of $\sim 5.4$ arcminutes from
the cluster center (note that SXC's field of view is $\sim 11\times 11$ arcmin). 

From Fig. 10 and from eq. 9 it follows that during a single 1 Msec
observation of the Perseus cluster with SXC, the turbulent broadening
at the level of a few hundred km/s can be measured in a central $1.8'
\times 1.8'$ pixel  and a map of the linewidth with the accuracy of the order of $10$ km/s
covering the area $\sim 11'\times 11'$ can be obtained. Given the expected
dependencies of the line broadening shown in Fig.6 and 7, such a dataset
would allow one to firmly establish the level of microturbulence in the
Perseus cluster and test the hypothesis that turbulent dissipation
acts as a mechanism for heating the ICM.

\section{Conclusions}
\label{sec:conclusions}

The dissipation of turbulent gas motions driven by AGN activity is a
plausible source of heat for the cooling gas in cluster cores. In this
case, we show that the expected width of the iron 6.7 keV line is well
above the thermal broadening for most plausible values of the turbulent
velocity. In our fiducial model of the Perseus cluster, the 6.7 keV
linewidth is larger than 10 eV.
Hence the proposed SXC micro-calorimeter is well suited for firmly establishing the level of
microturbulence in the brightest galaxy clusters.\\ 
As discussed in Sections 2.1 and 2.2 the linewidth is sensitive to both the 
radial dependence of the velocity amplitude and the 
"directionality" of the stochastic motions. This may complicate 
an unambiguous determination of the properties of the gas motions. However this degeneracy 
can be removed if the width of several lines, characteristic for 
different gas temperatures, can be measured. Assume for example that 
in addition to the 6.7 keV iron line, the width of the iron L-shell 
line is detected. This line is bright at lower temperatures, 
that are present only in the central region (cool core). Therefore the combined measurement
 of the width of two or more lines should be sufficient to discriminate among the simplest models with different "directionalities". 

\section*{Acknowledgments}
We acknowledge the support by the DFG grants CH389/3-2.  We would like
to thank Dan McCammon, Jan-Willem den Herder, Richard Kelley and Kazu
Mitsuda for the data and discussions about SXC. We wish to thank the anonimous referee for his/her useful comments. P.R. is grateful to
the International Max Planck Research School for its support, to
Andreas Bauswein, Marcus Br\"{u}ggen and Rasmus Voss for the helpful
discussions and to the "Velisti per Caso" of the Adriatica for their
enthusiasm.

\label{lastpage}
\end{document}